# Design of a vector multi-foci metalens for full stokes imaging polarimetry

*Shuyi Wang,[†] Tie Hu,[†] Shichuan Wang,[†] Ming Zhao,\* and Zhenyu Yang\**

Nanophotonics Laboratory, School of Optical and Electronic Information, Huazhong University of Science and Technology, Wuhan 430074, China

[†] These authors contributed equally to this work

\*Corresponding authors: zhaoming@hust.edu.cn, zyang@hust.edu.cn

Imaging polarimetry based on dielectric metasurface is well-known for its ultra-compactness and high integration. However, previous works suffer from low energy efficiency, limited restrictions on choice of target polarization states, or inability to focus light. Here, by inverse design, we numerically demonstrate a multi-foci metalens-based polarimetry that can simultaneously separate and focus the four free-chosen elliptical polarization states at the wavelength of 10.6 μm. Such a full-stokes polarimetry features an average absolute efficiency up to 54.63%, and an average relative error as low as 0.00137%. This spatial-multiplexing-free full stokes polarimetry exceeds the theoretical maximum efficiency of traditional polarization-filtering counterparts, and resolves the restriction faced by the orthogonal polarization-multiplexed method.

## 1. Introduction

Polarimetry, an optical system designed for the detection of polarization, has wide applications in optical communication,[1] biomedical imaging,[2] remote sensing,[3] and astronomy.[4] The Stokes vector,[5] which represents full polarization information of light, can be detected through at least four different polarization state intensity measurements.[3] Traditional polarimetry relies on bulky polarizers and retarders, and can mainly be categorized as "division of time (DoT)",[6] "division of amplitude (DoA)",[7] "division of focal plane (DoFP)",[8] and "division of aperture (DoA)".[9] These methods suffer from poor real-time performance, low integration, or crosstalk.

Metasurfaces, composed of subwavelength optical antenna array, have exhibited unprecedented ability to manipulate the phase, amplitude, and polarization. [10-14] Especially, owing to its ultracompactness, high integration, and multifunctionality, metasurface-based polarimetry has been widely studied.[15-19]. For example, a metasurface polarimetry is realized by spatially multiplexing polarization-sensitive metalenses.[20-22] By using the intrinsic birefringence of dielectric nanopillars,[23-25] a full-stokes imaging polarimetry based on three orthogonal polarization-multiplexing metasurfaces has been demonstrated, with an ideal maximum efficiency of 100%. Recently, a novel matrix Fourier optics polarimeter has been proposed,[26, 27] which splits any number of arbitrary target polarization state components with a single metasurface. some similar design strategies, such as super-cell,[28] unique diffraction patterns for polarization states,[29] adaptive target polarization selection,[30] are also proposed. However, there is still lack of a full-stokes polarimetry that simultaneously takes advantage of high efficiency, arbitrary selection of target polarization states, no spatial multiplexing, and light focusing.

In this work, by inversely optimizing the vector polarization responses of a single metasurface, we propose and demonstrate a compact full-stokes polarimeter with an absolute energy efficiency as high as 54.63% and an average relative error as low as 0.00137%. Different from the previous works[26, 27], the vector focal patterns at the target focal plane in this work are optimized and the proposed metasurface work as a combination of a vector polarization generator, optical splitter, and a focusing lens. Overcoming the limitations faced by the previous works,[20-27] our proposed multi-foci metalens-based polarimetry is realized by optimizing the response to four elliptical polarization states, which are freely selected and form a regular tetrahedron on a Poincaré sphere as an example.

## 2. Principle and design framework

**Figure 1**a displays the principle of the proposed multi-foci metalens-based polarimetry. When the incident light passes through the metasurface, four target polarization components marked by different polarization ellipses are extracted and focused on the center of four sub-squares divided by dashed lines on the focal plane, respectively. The target polarization states are shown in the Poincaré sphere in Figure 1b, which constructs the vertices of a regular tetrahedron to minimize the influence of noise.[3] Actually, there is no restriction on the selection of both the number and the type of target polarization states.

Figure 1c shows the optimization scheme of the metasurface. For incident plane wave whose polarization state is denoted by Jones vector $\boldsymbol{E}_{in}$, the output electric field $\boldsymbol{E}(\boldsymbol{r}_t)$ at any space point $\boldsymbol{r}_t$ can be obtained by using the equivalence principle and dyadic Green's function.[31-33] The calculation process is expressed as (see Section S1 for details of derivation)

$$\boldsymbol{E}(\boldsymbol{r}_t) = \mathbf{A}(\boldsymbol{r}_t)\boldsymbol{E}_{in} \tag{1}$$

where $\mathbf{A}(\boldsymbol{r}_t)$ is a matrix function distributed over the half space on one side of the metasurface, as is shown in Figure 1c. The explicit form of $\mathbf{A}(\boldsymbol{r}_t)$ is given by

$$\mathbf{A}(\boldsymbol{r}_t) = -2 \iint_S [\mathbf{G}_{ME}(\boldsymbol{r},\boldsymbol{r}_t) \cdot \boldsymbol{z} \times \mathbf{J}(\boldsymbol{r})] ds \tag{2}$$

where $\mathbf{G}_{ME}(\boldsymbol{r},\boldsymbol{r}_t)$ is the dyadic Green's function that represents the contribution of the unit magnetic source at $\boldsymbol{r}$ to the complex electric field at $\boldsymbol{r}_t$, $\boldsymbol{z}$ denotes the normal vector perpendicular to the metasurface, and $\mathbf{J}(\boldsymbol{r})$ is a Jones matrix function that characterizes the ability of metasurface to regulate light locally when taking into account the polarization response of metasurface.[26, 27] Equation 2 implies that $\mathbf{A}(\boldsymbol{r}_t)$ is a matrix fully determined by $\boldsymbol{r}_t$ and $\mathbf{J}(\boldsymbol{r})$. Such model is similar to the one proposed in "Matrix Fourier Optics", but here it cares about the spatial optical patterns rather than diffraction order.[26, 27]

Based on the model proposed above, an inverse design framework is built to optimize the multi-foci metalens, as is shown in Figure 1a. The key of the framework is to maximize the intensity at the target focus for target polarization states while minimizing those for their corresponding orthogonal ones. Here a single-objective optimization is summarized:

$$\begin{cases} \max\limits_{\mathbf{J}(\boldsymbol{r})} \quad \sum_l I_l \\ s.t. \quad \sum_l I_{o,l} + \text{std}(\{I_l\}) \leq D \end{cases} \tag{3}$$

Where $I_l$ and $I_{o,l}$ are the intensities of the target polarization states and their corresponding orthogonal polarization states at the specific points $\boldsymbol{r}_l$, respectively. std( ) and $D$ respectively denote standard deviation and a positive constant close to zero. In this optimization problem, $\sum_l I_l$ is maximized while every $I_{o,l}$ and the standard deviation of $\{I_l\}$ are close to zero, to ensure the polarization-sensitive behavior and the equivalent ability of metalens to deal with each target polarization state. The model proposed before gives $I_l = |\mathbf{A}(\boldsymbol{r}_l)\boldsymbol{E}_l|^2$ and $I_{o,l} = |\mathbf{A}(\boldsymbol{r}_l)\boldsymbol{E}_l^\perp|^2$, where $\boldsymbol{E}_l$ and $\boldsymbol{E}_l^\perp$ are Jones vectors of orthogonal polarization state pairs. In practice, an approximation is made that $\mathbf{J}(\boldsymbol{r})$ is constant within the region of each unit cell, whose Jones matrix is expressed as

$$\mathbf{J}(\phi_x, \phi_y, \theta) = \mathbf{R}(\theta) \begin{bmatrix} e^{i\phi_x} & 0 \\ 0 & e^{i\phi_y} \end{bmatrix} \mathbf{R}(-\theta) \tag{4}$$

where $\mathbf{R}(\theta)$ denotes the rotation matrix with the rotation angle $\theta$ relative to the $x$ axis, $\phi_x$ and $\phi_y$ are propagation phases along orthogonal axes. In this way, $\mathbf{J}(r)$ is discretized by a series of independent variables denoted by $(\phi_{x,i}, \phi_{y,i}, \theta_i)$, where $i$ denotes the indices of unit cells.

This optimization problem is solved by the inner-point method[29], implemented in the MATLAB optimization toolbox. Notably, a strategy for the selection of the initial solution is applied to fasten the optimization process. (See Section S2)

The principle of polarization detection based on the metasurface shown in Figure 1a is : Once the spot intensity at four points is got, the incident polarization state can be calculated by[3]

$$\mathbf{AS} = \mathbf{I} \tag{5}$$

where $\mathbf{S}$ denotes the Stokes vector of the incident light, $\mathbf{I} = [I_1 \ I_2 \ I_3 \ I_4]^T$ is a vector composed of spot intensity, and $\mathbf{A}$ is a $4 \times 4$ matrix required to be calibrated. The least-squares solution of $\mathbf{A}$ is given by

$$\mathbf{A} = \mathbf{IS}^T(\mathbf{SS}^T)^{-1} \tag{6}$$

where both $\mathbf{I} = [\cdots \mathbf{I}_n \cdots]$ and $\mathbf{S} = [\cdots \mathbf{S}_n \cdots]$ are matrices with 4 rows and $N$ columns, where $N$ denotes the number of polarization states used for calibration. After calibrating the matrix $\mathbf{A}$, stokes vector for any incident light can be reconstructed by using $\mathbf{S} = \mathbf{A}^{-1}\mathbf{I}$.

## 3. Results

To verify the above-mentioned method and principle, a multi-foci metalens polarimetry is designed, with a working wavelength of 10.6 μm, a size of 100 μm × 100 μm, and a working distance of 60 μm. As shown in **Figure 2**a, the meta-atom is composed of elliptical Ge nanopillar based on Ge square substrate, with a period $P$ of 2.5 μm and a height $H$ of 12 μm. The optical response of the unit cell can be obtained by a finite-difference time-domain (FDTD) approach. Figure 2b shows the phase shift and transmittance as a function of the major $D_x$ and minor axes $D_y$ of the elliptical nanopillars, under normal incidence of x-polarized light at the wavelength of 10.6 μm. When the $D_x$ and $D_y$ vary from 0.3 μm to 1 μm, a phase shift covers a range of $2\pi$ and transmittance exceeds 0.68. The top view of the constructed metasurface is displayed in Figure 2c.

**Figure 3**a depicts the slices of simulated intensity distribution along the propagation direction (**z**) under the incidence of four target polarization states, whose polarization ellipses are shown on the left side. The results show that there are always four spots corresponding to four target polarizations at every slice in the range of 35 μm and 65 μm along z-direction. And the optical power of the spot corresponding to incident target polarization is always higher than other spots. Also, the focal plane locates on *z* = 55 μm, where the maximum intensity is reached for each target polarization state. Therefore, the metasurface is verified with a multi-functionality to separate and focus different target polarization components. The normalized intensity distributions on the focal plane are shown in Figure 3b under the incidence of target polarization states (in brown box) and their corresponding orthogonal polarization states (in green box). Take the first row of Figure 3b as an example. The focus designed for the brown polarization ellipse has the maximum intensity while the other three are relatively weak in accordance with the case of regular tetrahedron distribution in the Poincaré sphere. While the orthogonal state is incident on the metasurface, the spot intensity of focus selected for the target polarization state is greatly suppressed, showing the great polarization sensibility to them. Here, the absolute energy efficiency is defined as the proportion of energy reaching the focal plane, and the simulated average value for the four target states is 54.63%, which surpasses the theoretical upper limit of traditional counterparts.[23]. **Figure 4**a shows both the optimization(blue) and simulation(red) results of the contrast for four target polarization states on their corresponding sub-squares. The simulated results are in good agreement with the design values, and it is noted that the simulated contrasts are lower because of the approximation made in the model and the error from the sifting process. Here, the $Contrast = \frac{P - P_\perp}{P + P_\perp}$, where $P$ and $P_\perp$ denote the optical power within one sub-square under the incidence of the target and orthogonal polarizations, respectively. The optical power is calculated by summarizing the values in one sub-square and is divided by the pixel number of the whole image.

The results of polarization detections are represented on the Poincaré sphere in Figure 4b, giving some examples of polarization detection of incident polarization states. There is a comparison between the original (blue circle) and restored(red dot) states, and for each pair, the dot almost coincides with the center of the circle. The detailed relative error for each polarization state is shown in Table S1, and the simulated average relative error is 0.00137%.

## 4. Conclusion

In summary, an inverse design framework is proposed to optimize the optical vector responses of spatial-multiplexing-free multi-foci metalens, with the ability to simultaneously extract and focus the four target polarization states from the incident light. Then, a full-stokes polarimetry working at the wavelength of 10.6 μm has been built, featuring an average absolute energy efficiency up to 54.63%, and an average relative error as low as 0.00137%. In addition, there is no restriction on the selection of target polarization states and distribution of regular tetrahedrons in Poincaré sphere has been chosen specifically to lower the influence of noise. Moreover, broadband full-stokes polarimetry may be achieved when dispersion engineering is introduced into our design.

**Funding**

National Natural Science Foundation of China (62075073, 62075129, 62135004); Department of Science and Technology, Hubei Provincial People's Government (2021BAA003); Science and Technology Foundation of State Grid Corporation of China (52180023000M): Research on industrial video image enhancement and recognition technology of power grid in all-weather environment based on multi-dimensional optical parameters fusion and pulse computing.

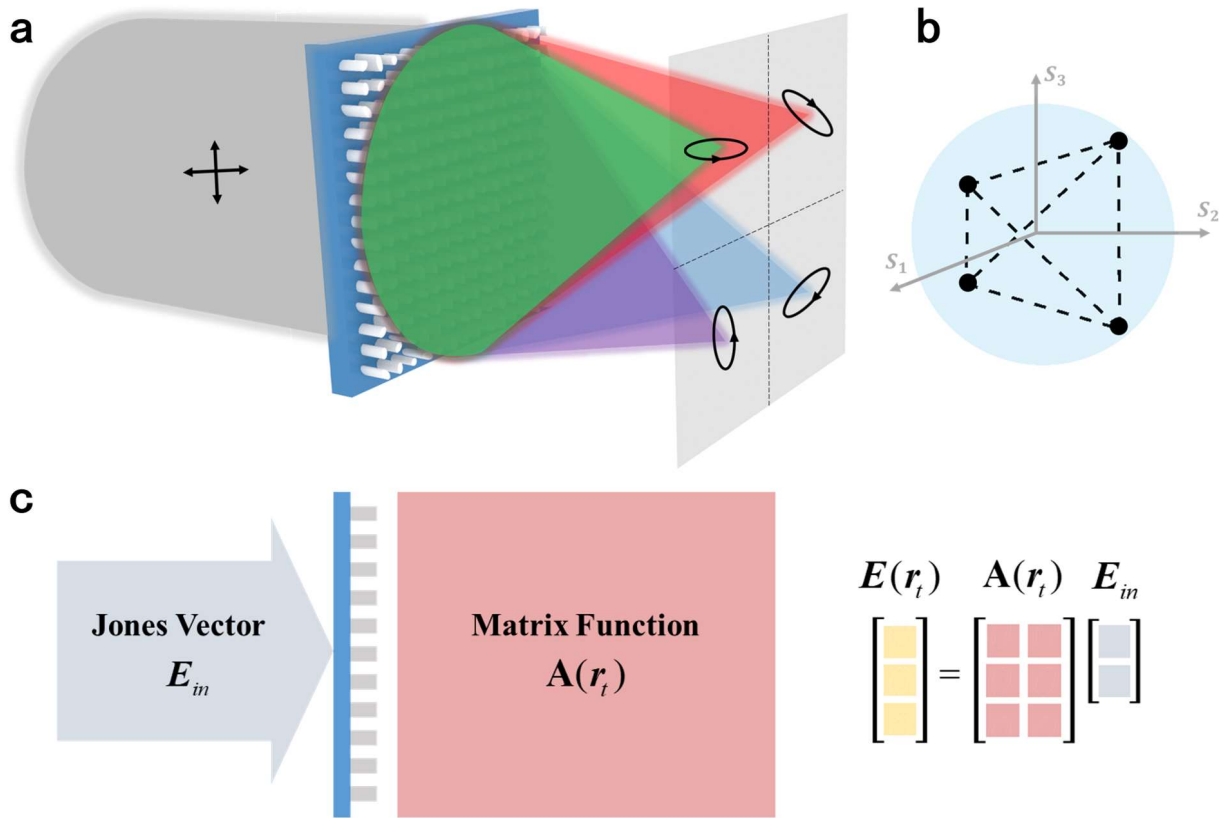

**Figure 1.** Principle of the vector multi-foci metalens-based polarimetry. a) Scheme of the multi-foci metalens designed in this work. Different colors represent different polarization components of interest, marked by different polarization ellipses. b) Target polarization states forming a regular tetrahedron in the Poincaré sphere. c) Scheme of the forward solving model for metasurface.

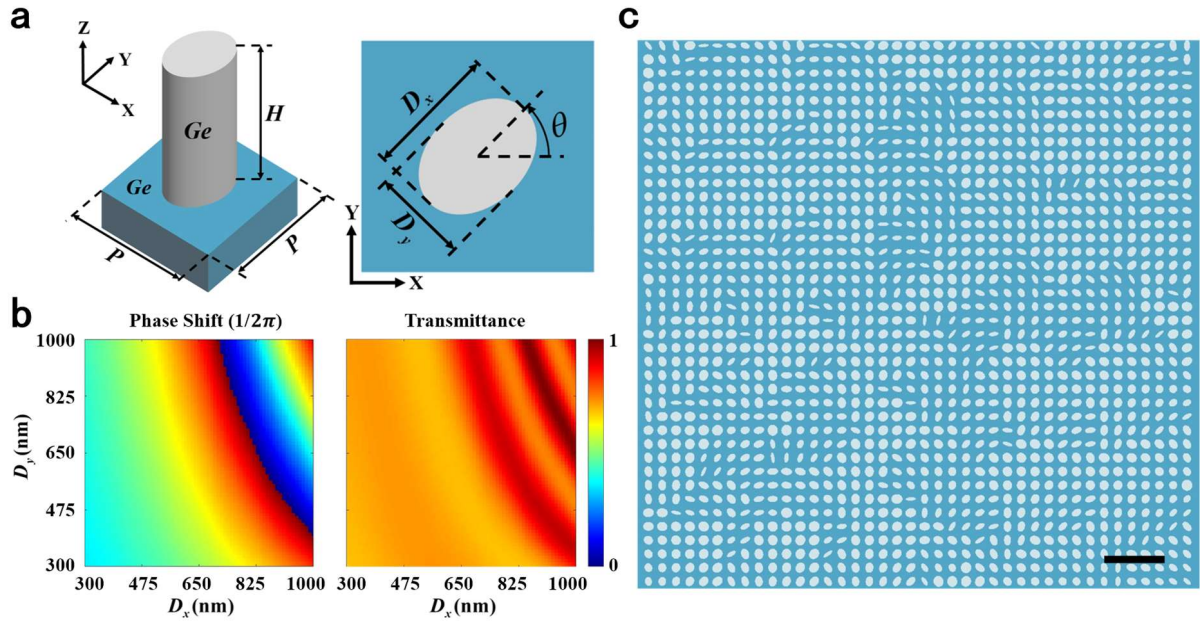

**Figure 2.** Design of the metasurface. a) Structure of meta-atom composed of Ge elliptical nanopillar on Ge substrate, with a pillar height $H$ of 12 µm and Period $P$ of 2.5 µm. b) Phase shift and transmittance with a function of major axis $D_x$ and minor axis $D_y$ under normal incidence of x-polarized light. c) Top view of the constructed metasurface. Scale bar is 10 µm.

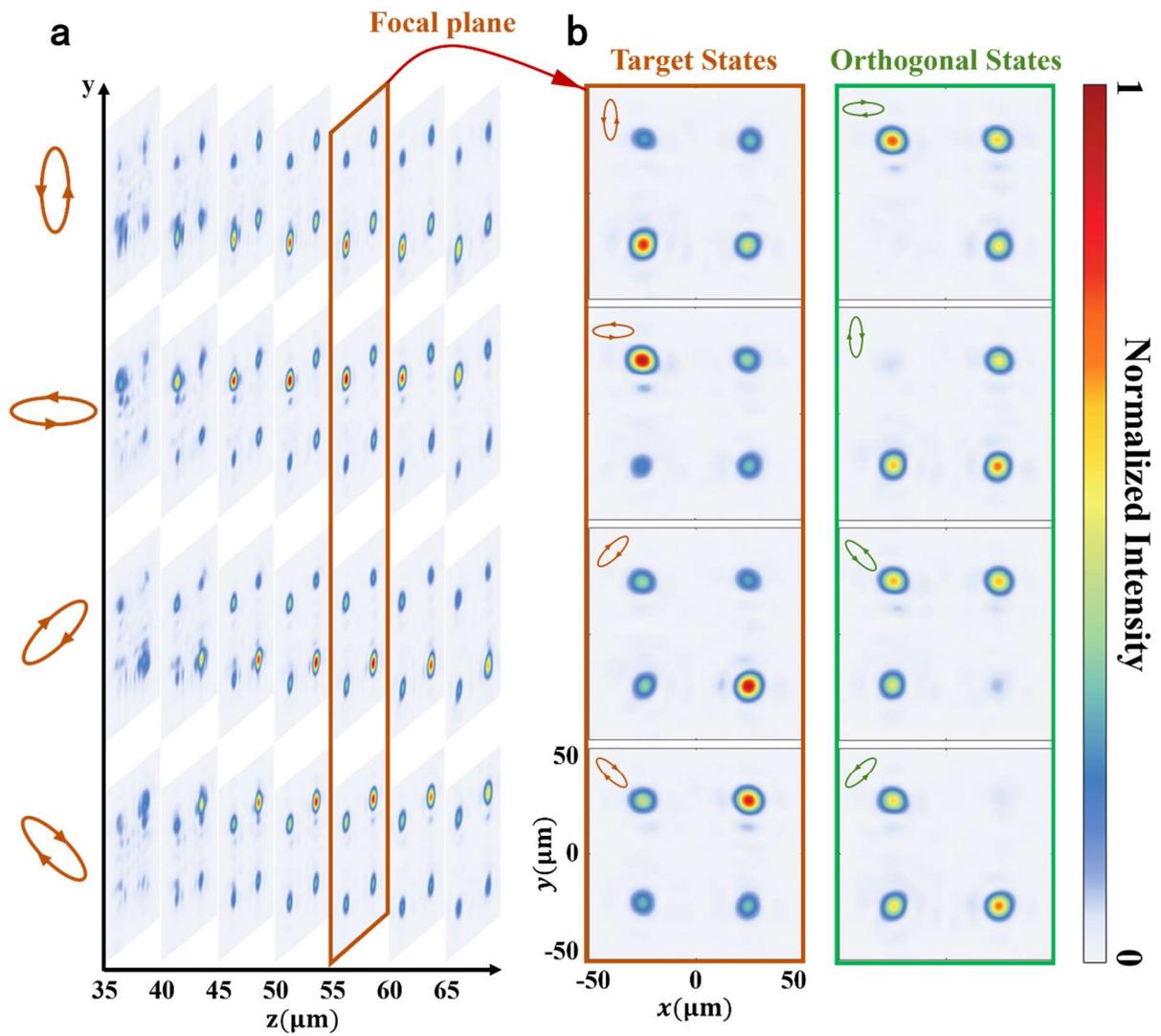

**Figure 3.** Optical responses of the metasurface. a) Slides of intensity distribution along z axis under the incidence of target polarization states, labeled by polarization ellipse in brown. b) Normalized intensity distributions on the focal plane when target polarization states(brown) or orthogonal polarization states(green) are incident on the metasurface.

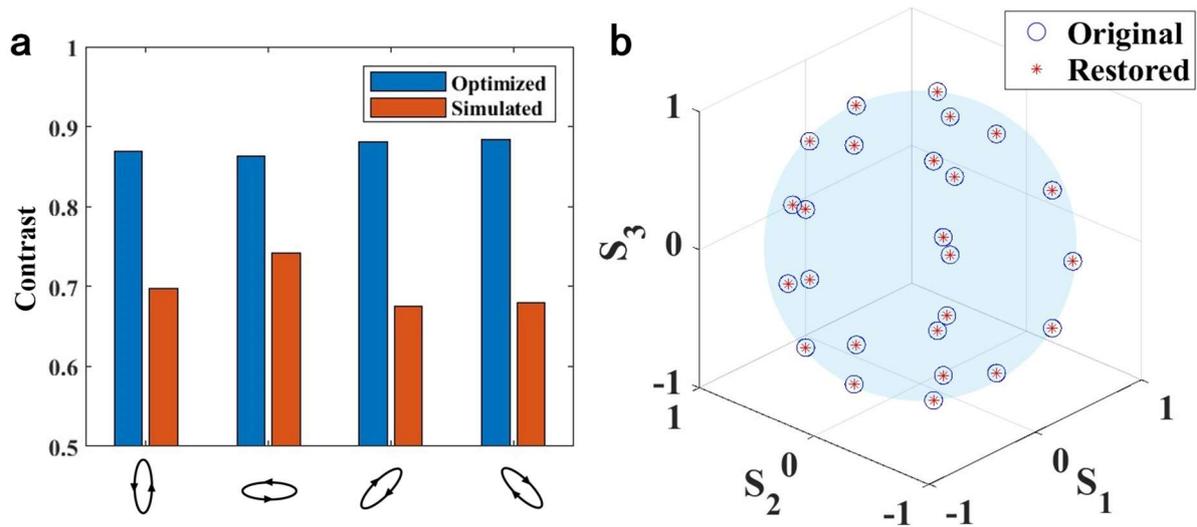

**Figure 4.** Polarization reconstruction result of the polarimetry. a) Optimized (blue) and simulated (red) contrast for four target polarization states. b) Original and restored results of several polarization states shown in Poincaré sphere.